\documentclass[%
 reprint,
superscriptaddress,
nofootinbib,
amsmath,amssymb,
 aps,
 prl,
floatfix,
longbibliography
]{revtex4-2}
\usepackage[colorlinks=true,citecolor=blue,filecolor=blue,linkcolor=blue,urlcolor=blue,pdftex]{hyperref}
\usepackage{orcidlink} 
\usepackage{currfile} 
\usepackage{graphicx}
\usepackage{dcolumn}
\usepackage{bm}
\usepackage{siunitx}
\usepackage{booktabs}
\usepackage{xspace}
\usepackage[utf8]{inputenc}
\usepackage{booktabs,mathrsfs,verbatim,amsmath,soul}
\usepackage{xspace} 
\usepackage{upgreek}
\usepackage{textcomp}
\usepackage{gensymb}
\usepackage{tabularx}
\usepackage{xcolor}
\usepackage{multirow}
\usepackage{tabularx}
\usepackage{orcidlink} 
\usepackage{currfile} 

\usepackage[mathlines]{lineno}

\usepackage[separate-uncertainty=true]{siunitx}
\DeclareSIUnit\gevm{\GeV\per\clight\squared}

\DeclareSIUnit{\events}{events}
\DeclareSIUnit{\ton}{t}
\DeclareSIUnit{\tonwhole}{tonne}
\DeclareSIUnit{\year}{y}
\DeclareSIUnit{\yearwhole}{year}
\DeclareSIUnit{\tonyear}{t\cdot y}
\DeclareSIUnit{\eventspertonyear}{\events\per\ton\per\year}
\DeclareSIUnit{\eventspertonyearwhole}{\events\per\tonwhole\per\yearwhole}

\newcommand{\itsec}[1]{{\it#1}---}
\newcommand{\cevns}{{\text{CE}\ensuremath{\nu}\text{NS}}\xspace}
\newcommand{\beight}{\ensuremath{{}^8\mathrm{B} }\xspace}
\newcommand{\gevcsq}{\ensuremath{\mathrm{GeV}/c^2}}

\newcommand{\gevm}{\gevcsq}

\newcommand{\keVnr}{\ensuremath{\mathrm{keV_\mathrm{nr}}}\xspace}
\newcommand{\keVer}{\ensuremath{\mathrm{keV_\mathrm{ee}}}\xspace}

\begin{document}

\preprint{APS/123-QED}
\title{Light Dark Matter Search with 7.8 Tonne-Year of Ionization-Only Data in XENONnT}


\newcommand{\bologna}{\affiliation{Department of Physics and Astronomy, University of Bologna and INFN-Bologna, 40126 Bologna, Italy}}
\newcommand{\chicago}{\affiliation{Department of Physics, Enrico Fermi Institute \& Kavli Institute for Cosmological Physics, University of Chicago, Chicago, IL 60637, USA}}
\newcommand{\coimbra}{\affiliation{LIBPhys, Department of Physics, University of Coimbra, 3004-516 Coimbra, Portugal}}
\newcommand{\columbia}{\affiliation{Physics Department, Columbia University, New York, NY 10027, USA}}
\newcommand{\lngs}{\affiliation{INFN-Laboratori Nazionali del Gran Sasso and Gran Sasso Science Institute, 67100 L'Aquila, Italy}}
\newcommand{\mainz}{\affiliation{Institut f\"ur Physik \& Exzellenzcluster PRISMA$^{+}$, Johannes Gutenberg-Universit\"at Mainz, 55099 Mainz, Germany}}
\newcommand{\mpik}{\affiliation{Max-Planck-Institut f\"ur Kernphysik, 69117 Heidelberg, Germany}}
\newcommand{\munster}{\affiliation{Institut f\"ur Kernphysik, University of M\"unster, 48149 M\"unster, Germany}}
\newcommand{\nikhef}{\affiliation{Nikhef and the University of Amsterdam, Science Park, 1098XG Amsterdam, Netherlands}}
\newcommand{\nyuad}{\affiliation{New York University Abu Dhabi - Center for Astro, Particle and Planetary Physics, Abu Dhabi, United Arab Emirates}}
\newcommand{\purdue}{\affiliation{Department of Physics and Astronomy, Purdue University, West Lafayette, IN 47907, USA}}
\newcommand{\rice}{\affiliation{Department of Physics and Astronomy, Rice University, Houston, TX 77005, USA}}
\newcommand{\stockholm}{\affiliation{Oskar Klein Centre, Department of Physics, Stockholm University, AlbaNova, Stockholm SE-10691, Sweden}}
\newcommand{\subatech}{\affiliation{SUBATECH, IMT Atlantique, CNRS/IN2P3, Nantes Universit\'e, Nantes 44307, France}}
\newcommand{\torino}{\affiliation{INAF-Astrophysical Observatory of Torino, Department of Physics, University  of  Torino and  INFN-Torino,  10125  Torino,  Italy}}
\newcommand{\ucsd}{\affiliation{Department of Physics, University of California San Diego, La Jolla, CA 92093, USA}}
\newcommand{\wis}{\affiliation{Department of Particle Physics and Astrophysics, Weizmann Institute of Science, Rehovot 7610001, Israel}}
\newcommand{\zurich}{\affiliation{Physik-Institut, University of Z\"urich, 8057  Z\"urich, Switzerland}}
\newcommand{\paris}{\affiliation{LPNHE, Sorbonne Universit\'{e}, CNRS/IN2P3, 75005 Paris, France}}
\newcommand{\freiburg}{\affiliation{Physikalisches Institut, Universit\"at Freiburg, 79104 Freiburg, Germany}}
\newcommand{\napels}{\affiliation{Department of Physics ``Ettore Pancini'', University of Napoli and INFN-Napoli, 80126 Napoli, Italy}}
\newcommand{\nagoya}{\affiliation{Kobayashi-Maskawa Institute for the Origin of Particles and the Universe, and Institute for Space-Earth Environmental Research, Nagoya University, Furo-cho, Chikusa-ku, Nagoya, Aichi 464-8602, Japan}}
\newcommand{\laquila}{\affiliation{Department of Physics and Chemistry, University of L'Aquila, 67100 L'Aquila, Italy}}
\newcommand{\tokyo}{\affiliation{Kamioka Observatory, Institute for Cosmic Ray Research, and Kavli Institute for the Physics and Mathematics of the Universe (WPI), University of Tokyo, Higashi-Mozumi, Kamioka, Hida, Gifu 506-1205, Japan}}
\newcommand{\kobe}{\affiliation{Department of Physics, Kobe University, Kobe, Hyogo 657-8501, Japan}}
\newcommand{\kit}{\affiliation{Institute for Astroparticle Physics, Karlsruhe Institute of Technology, 76021 Karlsruhe, Germany}}
\newcommand{\tsinghua}{\affiliation{Department of Physics \& Center for High Energy Physics, Tsinghua University, Beijing 100084, P.R. China}}
\newcommand{\ferrara}{\affiliation{INFN-Ferrara and Dip. di Fisica e Scienze della Terra, Universit\`a di Ferrara, 44122 Ferrara, Italy}}
\newcommand{\groningen}{\affiliation{Nikhef and the University of Groningen, Van Swinderen Institute, 9747AG Groningen, Netherlands}}
\newcommand{\westlake}{\affiliation{Department of Physics, School of Science, Westlake University, Hangzhou 310030, P.R. China}}
\newcommand{\shenzhen}{\affiliation{School of Science and Engineering, The Chinese University of Hong Kong (Shenzhen), Shenzhen, Guangdong, 518172, P.R. China}}
\newcommand{\coimbrapoli}{\affiliation{Coimbra Polytechnic - ISEC, 3030-199 Coimbra, Portugal}}
\newcommand{\heidelberg}{\affiliation{Kirchhoff-Institute for Physics, Heidelberg University, 69120 Heidelberg, Germany}}
\newcommand{\roma}{\affiliation{INFN-Roma Tre, 00146 Roma, Italy}}
\newcommand{\bucknell}{\affiliation{Department of Physics \& Astronomy, Bucknell University, Lewisburg, PA, USA}}



\author{E.~Aprile\,\orcidlink{0000-0001-6595-7098}}\columbia
\author{J.~Aalbers\,\orcidlink{0000-0003-0030-0030}}\groningen
\author{K.~Abe\,\orcidlink{0009-0000-9620-788X}}\tokyo
\author{M.~Adrover\,\orcidlink{0123-4567-8901-2345}}\zurich
\author{S.~Ahmed Maouloud\,\orcidlink{0000-0002-0844-4576}}\paris
\author{L.~Althueser\,\orcidlink{0000-0002-5468-4298}}\munster
\author{B.~Andrieu\,\orcidlink{0009-0002-6485-4163}}\paris
\author{E.~Angelino\,\orcidlink{0000-0002-6695-4355}}\lngs
\author{D.~Ant\'on~Martin\,\orcidlink{0000-0001-7725-5552}}\chicago
\author{S.~R.~Armbruster\,\orcidlink{0009-0009-6440-1210}}\mpik
% ...
\collaboration{XENON Collaboration}\email[]{xenon@lngs.infn.it}\noaffiliation

\date{\today}

\begin{abstract}We report on a blinded search for dark matter (DM) using ionization-only (S2-only) signals in XENONnT with a total exposure of $7.83\,\mathrm{tonne}\times\mathrm{year}$ over 579 days in three science runs.
Dedicated background suppression techniques and the first complete S2-only background model in XENONnT provide sensitivity to nuclear recoils of [0.5, 5.0] $\mathrm{keV_\mathrm{nr}}$ and electronic recoils of [0.04, 0.7] $\mathrm{keV_\mathrm{ee}}$.
No significant excess over the expected background is observed, and we set 90\% confidence level upper limits on spin-independent DM--nucleon and spin-dependent DM--neutron scattering for DM masses between 3 and 8 $\mathrm{GeV}/c^2$, as well as on DM–electron scattering, axion-like particles, and dark photons, improving on previous constraints.
For spin-independent DM--nucleon scattering, we exclude cross sections above $6.0\times10^{-45}$\,cm$^2$ at a DM mass of 5 $\mathrm{GeV}/c^2$, pushing the XENONnT sensitivity closer to the region where coherent elastic neutrino-nucleus scattering ($\text{CE}\nu\text{NS}$) becomes an irreducible background.

\end{abstract}

\maketitle

\itsec{Introduction}The XENONnT experiment \cite{XENON:2024wpa}, operating at INFN Laboratori Nazionali del Gran Sasso, is primarily designed to search for weakly interacting massive particles (WIMPs) \cite{Roszkowski:2017nbc,Jungman:1995df,Bertone:2016nfn,XENON:2025vwd} with a mass scale of \gevcsq --{\ensuremath{\mathrm{TeV}/c^2}}.
Using events that produce few scintillation photons and ionization electrons, we previously conducted a search for light dark matter (DM) in the neutrino fog from solar \beight coherent elastic neutrino--nucleus scattering (\cevns) \cite{XENON:2024hup, XENON:2024ijk}.
A complementary search that uses only the ionization signal (S2-only) \cite{XENON:2019gfn,XENON:2021qze,XENON:2024znc,Zhang:2025ajc,PandaX:2022xqx} remains robust when scintillation becomes too weak, extending the sensitivity of XENONnT to even lower recoil energies and a broader class of light DM models.
The previous S2-only analysis in XENON1T~\cite{XENON:2019gfn} demonstrated the potential of the S2-only approach but was limited by an incomplete background model.
In this Letter, we report on the first S2-only search for light DM in XENONnT employing a full background model.

The XENONnT dual-phase time projection chamber (TPC) contains 5.9 tonnes of liquid xenon. 
Particle interactions with xenon atoms produce prompt scintillation light (S1) and ionization electrons \cite{Aprile:2009dv}.
The electrons drift upward under an electric field of $\sim$\SI{23}{\volt/\centi\meter} between the cathode and gate electrodes.
Perpendicular wires across the gate and anode electrodes provide mechanical support.
Along the drift path, attachment to electronegative impurities causes an exponential loss of electrons characterized by the ''electron lifetime'' \cite{Plante:2022khm}, and surviving electrons are extracted into the gas phase, where they generate proportional scintillation (S2) \cite{Monteiro:2007vz}.
The maximum drift time for electrons drifting from cathode to gate is $\sim$\SI{2.2}{ms}.
Both signals are recorded by photomultiplier tubes (PMTs) at the top and bottom of the TPC and converted into photoelectrons (PE).
S2s are composed of single electron (SE) signals; the SE gain is defined as the mean number of PE per electron.
Event $(x,y)$ positions are reconstructed from the S2 light pattern on the top PMTs, and $z$ from the time difference between S1 and S2.
Particle interactions may produce nuclear recoils (NR) with xenon nuclei or electronic recoils (ER) with atomic electrons; the two populations are distinguished by their different response in the S1–S2 signal size \cite{XENON:2024xgd,Szydagis:2022ikv}.
Further details on XENONnT are provided in~\cite{XENON:2024wpa}.

\itsec{Dataset}This DM search combines data from three XENONnT science runs (SRs), denoted SR0, SR1, and SR2, with livetime of 110.2, 177.8, and 291.5 days, respectively.
The data were collected between May 2021 and March 2025, and the total livetime is 579.5 days after accounting for data acquisition dead time \cite{XENON:2022vye}.
Due to the presence of photoionizable impurities \cite{XENON100:2013wdu} introduced in SR1, additional purifiers were installed in the radon removal system mid-SR2, allowing their removal thereafter. The total background is dominated by instrumental components, arising from detector-induced electron and photon losses~\cite{LZ:2026hpq} and SE activity~\cite{XENON:2021qze,LZ:2025mmr}, rather than genuine low-energy recoils in the liquid xenon bulk. To mitigate this, we exclude the 5\% of runs with the highest SE rates. The typical values of the detector response parameters, including photon gain $g_1$, charge gain $g_2$, SE gain, and electron lifetime, can be found in \cite{XENONCollaboration:2024bil,XENON:2025vwd}.

\begin{figure}[t!]
    \centering
    \includegraphics[width=1.0\linewidth]{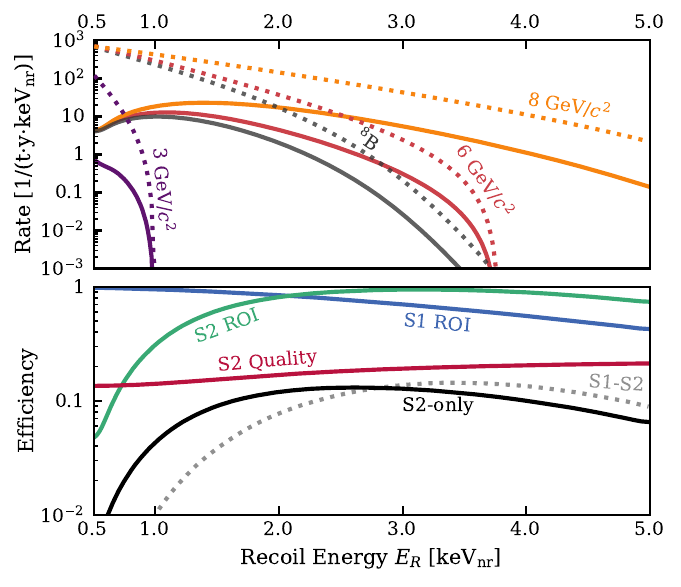}
    \caption{\label{fig:efficiency}
    Top: The SI DM-nucleon spectra for 3, 6, and 8~\gevcsq~masses with a cross section of $4.4\times10^{-45}$ cm$^2$, shown before (dotted) and after (solid) applying the total signal detection efficiency. The solar $^8$B CE$\nu$NS spectrum (gray) is shown for comparison.
    Bottom: The S2-only signal efficiency (black) is the product of the individual efficiencies for the S1 region of interest (ROI) (blue), S2 ROI (green), and S2 quality selection (red), compared to the S1-S2 analysis efficiency \cite{XENON:2024hup} (gray dashed). All efficiencies shown are exposure-weighted averages over SR0, SR1, and SR2.
    }
\end{figure}

\itsec{Signal}We consider several DM models producing NR or ER energies in the regions of interest (ROIs) of [0.5, 5]~\keVnr and [0.04, 0.7]\,\keVer, respectively. 
For the NR signals, we study spin-independent (SI) DM--nucleon \cite{Lewin:1995rx} and spin-dependent (SD) DM--nucleon \cite{Menendez:2012tm} scattering for DM masses between 3 and 8~\gevcsq, assuming the astrophysical parameters in~\cite{Baxter:2021pqo}.
The DM--nucleon recoil energy spectra for several benchmark DM masses are shown in Fig.~\ref{fig:efficiency}. 
For the SD scattering, we adopt the nuclear form factor from~\cite{Hoferichter:2020osn}.
We consider DM--nucleon interactions with cross section $\sigma$ and momentum transfer $q$ with a mediator mass $m_\phi$ \cite{DelNobile:2015uua, Fornengo:2011sz}, in which the differential rate factor $\sigma m_\phi^4/(q^2/c^2+m_\phi^2)^2$ scales with $\sigma m_\phi{ }^4c^4/q^4$ in the $m_\phi \ll q_0/c$ limit.
For the ER signals, we consider DM scattering off xenon bounded electrons.
The dark matter--electron scattering rates are computed using atomic ionization factors from~\cite{Essig:2012yx,Essig:2017kqs}, including both the primary electron and shell de-excitation photon energies with conservative assumptions for de-excitation electrons the same as \cite{XENON:2019gfn}; as a comparison, we also evaluate the rates based on the total deposited energy using the ionization factor from~\cite{Caddell:2023zsw}.
We compute the DM recoil energy spectrum using \cite{jelle_aalbers_2023_7636982}.
Bosonic DM candidates such as axion-like particles and dark photons of mass $m_\chi$ can be absorbed by xenon atoms, producing monoenergetic ER signals at $E_\chi=m_\chi c^2$ \cite{Pospelov:2008jk}. We follow models detailed in~\cite{Bloch:2016sjj} and use the xenon photoelectric cross section from~\cite{Henke:1993eda}.

We model the low-energy NR yield using a $^{88}$Y/Be neutron source~\cite{XENON:2024kbh,collar_applications_2013}. The NR yield below 0.5~\keVnr is set to zero to remain conservative. A combined $^{37}$Ar and $^{220}$Rn calibration dataset is fit jointly to determine the ER response~\cite{XENON:2024xgd}.
To generate realistic S2 waveforms and PMT patterns, we use the simulation tool \textsc{fuse} \cite{xenon-fuse} equipped with data-driven electron diffusion constant and optical response maps matched to $^{83\mathrm{m}}$Kr calibration data.
To capture the “ambience” of an uncorrelated DM signal, including the surrounding high-energy events and electron emission activities, we randomly inject (salt) simulated DM S2 signals into the real data stream and process them through the reconstruction chain \cite{strax, straxen}.
This procedure allows for a full simulation-driven signal efficiency evaluation.

\itsec{Background origin and data selection}Events are selected within the S2 area 100 (120) to 500~PE in SR0/SR1 (SR2), corresponding to $\sim$3 (4) to $\sim$16 extracted electrons.
The S2 threshold is set to suppress instrumental backgrounds and is higher in SR2 due to an update of the low-level processing to mitigate SE pile-ups.
Events without an S1 or with an S1 area below 3 PE are accepted to retain events accidentally paired with small S1s, while events with S1 area above 3 PE are rejected.
The analysis dimension is the corrected S2 ($\mathrm{cS2}$) \cite{xenonnt:analysis1} in the ROI of [80, 500] PE, after the time- and spatial-dependent SE gain and extraction efficiency corrections \cite{XENONCollaboration:2024bil}.

The cathode background arises from radioactive decays of $^{210}$Pb and its daughters from the $^{222}$Rn decay chain \cite{XENON:2017ciq}. 
The S1 from electron-ion recombination is suppressed by the strong electric field near the cathode~\cite{Szydagis:2022ikv}, while part of the ionization electrons are lost along downward-pointing field lines below the cathode wires~\cite{Linehan:2022hdb,LZ:2026hpq}.
The longer drift time leads to larger S2 width compared to events in the TPC bulk through longitudinal diffusion \cite{Sorensen:2011qs}.
To maximize its rejection power, we construct a gradient-boosted decision tree (BDT) using the S2 area together with multiple S2 waveform features. The BDT is trained on simulated signal and cathode datasets to improve classification accuracy. An optimized threshold on the BDT score rejects about $70\%$ of cathode backgrounds while keeping approximately $80\%$ of the signal.
In addition, a complementary drift time veto removes events that exhibit S1-like signals within one maximum drift time prior to the S2, rejecting roughly $14\%$ of cathode events with $\sim 0.5\%$ signal loss.

\begin{figure}[h!t]
    \centering
    \includegraphics[width=1\linewidth]{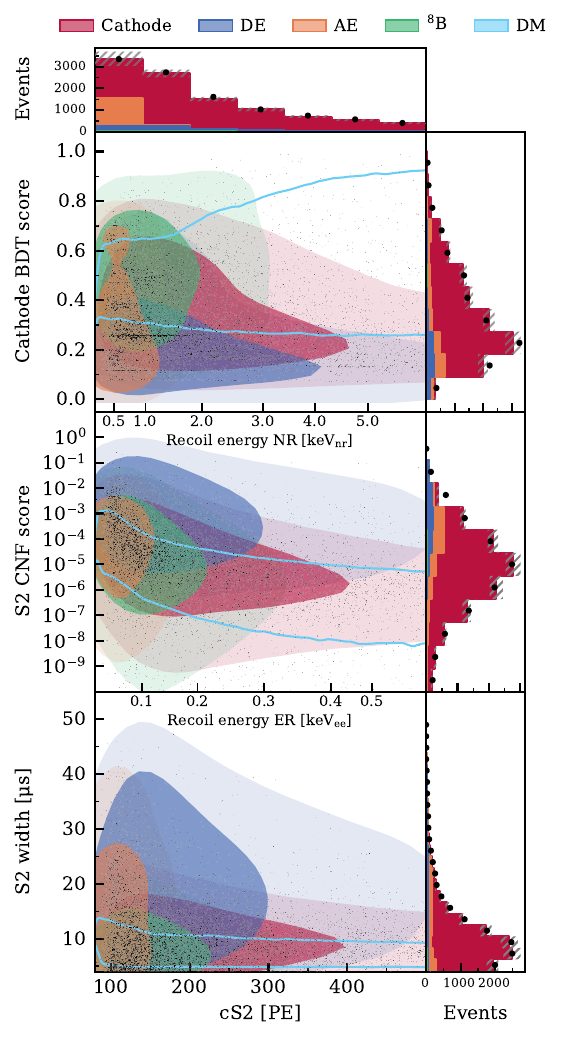}
    \caption{\label{fig:background_origin}
    Comparison of observed events (black dots) with the expected background components in the enlarged ROI of all science runs.
    Colored contours show the modeled distributions of cathode (red), delayed electrons (blue), accidental electrons (orange), and \beight~\cevns (green), with dark and light shades representing the $1\sigma$ and $2\sigma$ contours.
    A signal distribution (cyan) $1\sigma$ contour is shown, assuming a flat spectrum of signal recoil energy.
    The validation is performed in four dimensions: cS2 (energy scale), S2 CNF score (ambience), cathode BDT score and S2 width (waveform).
    Projections along each axis are shown as top and side histograms; shaded bands indicate background model statistical and systematic uncertainties.
    }
\end{figure}

Large S2 signals from energetic events, such as $\beta$ and $\gamma$ from the detector material \cite{XENON:2021mrg}, produce isolated S2s extending to $\sim$\SI{1}{\second} after the large S2 \cite{XENON:2021qze, LZ:2025mmr, LUX:2020vbj}.
Isolated S2s within \SI{2.2}{ms} are dominated by photoionizable impurities in the bulk or surfaces \cite{XENON100:2013wdu}; beyond the maximum drift time, they are referred to as delayed electrons (DE) and the rate approximately follows a power-law decay.
Selections developed for the S1--S2 analysis in~\cite{XENON:2024hup, XENON:2024ijk} are applied to suppress DE.
However, the DE rate and its distributions in charge, waveform, temporal and spatial observables depend nontrivially on the energy and location of the preceding large S2s and on run-dependent detector conditions \cite{XENON:2021qze, LZ:2025mmr, LUX:2020vbj}.
To achieve optimized DE modeling and mitigation, we employ a conditional normalizing flow (CNF) \cite{Papamakarios:2019fms} implemented in \textsc{PyTorch} with the \textsc{nflows} library \cite{paszke2019_pytorch, nflows} trained on science data to learn the joint probability $P(N, \mathbf{x} \mid \mathbf{y}) = \text{Poisson}(N \mid \lambda(\mathbf{y})) \times  P(\mathbf{x} \mid \mathbf{y})$ per SR,
where $N$ ($\lambda(\mathbf{y})$) is the observed (expected) number of isolated S2s in a \SI{1}{\second} window, $\mathbf{y}$ encodes run conditions and the preceding large S2 size and positions, and $\mathbf{x}$ describes the DE S2s, including the delayed time, size, position, waveform, and optical patterns.
For each DM candidate, the CNF model evaluates its correlation with all preceding large S2s within \SI{1}{\second} and assigns the maximum conditional likelihood as the CNF score.
A dedicated DE BDT combines the CNF score and waveform features, rejecting $\sim$95\% of the remaining DE background with $\sim$90\% signal efficiency.

The pile-up of randomly emitted SE signals can be reconstructed as a single S2, which constitutes the accidental electron (AE) background.
Its key distinction from DE is that AE includes a component that cannot be mitigated using the CNF score, hypothesized to originate from constant SE emission.
We require each DM candidate to be temporally isolated, with no other S2 in the \SI{2.2}{ms} preceding the candidate S2.
AEs combine SEs from different locations, producing unphysical S2 patterns on the top PMTs.
A data-driven S2 pattern likelihood selection, built from single-site $^{83\mathrm{m}}$Kr calibration data using a \textsc{TensorFlow}-based \cite{tensorflow2015-whitepaper} multilayer perceptron model, rejects $\sim$75\% of AE events while retaining $\sim$99\% signal efficiency; the remaining background arises from SE pile-up at indistinguishable locations.

Surface backgrounds from $^{222}$Rn progeny on TPC walls are suppressed by fiducialization in $r$ ($\sqrt{x^2+y^2}$) dimension.
Charge-up on the wall distorts electron drift paths, causing reconstructed $r$ to vary with $z$ and requiring a $z$ reconstruction to correct field effects \cite{XENONCollaboration:2024bil}.
We developed a BDT regression model to reconstruct the S2-only event $z$ trained on simulation using S2 area and S2 waveform.
After correcting $r$ for field effects, events with $r<$\SI{60.15}{cm} (SR0) and $r<$\SI{59.60}{cm} (SR1, SR2) are selected, yielding fiducial masses of 4.98, 5.01, and 4.87 tonnes, respectively,  as evaluated with signal salting and $^{37}$Ar calibration data \cite{XENON:2022ivg}.
The fiducial mass differences among SRs arise from liquid xenon density variations and field distortion correction, for which we assign a $\sim$5\% systematic uncertainty based on different evaluation methods.
Events near perpendicular wires \cite{XENON:2024wpa} are removed due to electrode SE emission.
S1–S2 events reconstructed in the cathode ($[-152.5,-146.5]$ cm) and gas ($[-4.5,0]$ cm) regions are excluded.
Surface backgrounds become negligible after fiducialization.

\begin{table*}[htbp]
\caption{\label{tab:table1} 
Expected and best-fit event numbers of each background component and a $6~\mathrm{GeV}/c^{2}$ DM assuming the DM--nucleon SI interactions in three science runs for the science ROI are listed.
The best-fit DM event rate is zero.
Only systematic uncertainties are included in the quoted expectations.
} 
\begin{ruledtabular}
\begin{tabular}{
  @{}l
  r@{~}l
  r@{~}l
  r@{~}l
  @{}
}
 Science run (exposure)
 & \multicolumn{2}{c}{SR0 (\SI{1.50}{\tonyear})} 
 & \multicolumn{2}{c}{SR1 (\SI{2.44}{\tonyear})} 
 & \multicolumn{2}{c}{SR2 (\SI{3.89}{\tonyear})} \\[3pt]
 Component
 & \multicolumn{1}{c}{Expectation} & \multicolumn{1}{c}{Best fit} 
 & \multicolumn{1}{c}{Expectation} & \multicolumn{1}{c}{Best fit} 
 & \multicolumn{1}{c}{Expectation} & \multicolumn{1}{c}{Best fit} \\
\midrule
Cathode 
 & $480 \pm 70$ & $477^{+25}_{-24}$ 
 & $660 \pm 70$ & $726^{+28}_{-27}$ 
 & $1210 \pm 90$ & $1080 \pm 30$ \\
Delayed electron 
 & $1.3 \pm 0.5$ & $1.3 \pm 0.5$ 
 & $0.34 \pm 0.07$ & $0.34 \pm 0.07$ 
 & $17.2 \pm 2.3$ & $17.1 \pm 2.3$ \\
Accidental electron 
 & $97 \pm 17$ & $89 \pm 13$ 
 & $108 \pm 8$ & $106 \pm 8$ 
 & \multicolumn{2}{c}{--} \\
$^8$B CE$\nu$NS 
 & $21 \pm 5$ & $18^{+5}_{-4}$ 
 & $29 \pm 7$ & $26^{+7}_{-6}$ 
 & $32.3 \pm 8$ & $29^{+8}_{-6}$ \\
\midrule
Total background 
 & $600 \pm 80$ & $586 \pm 28$ 
 & $800 \pm 80$ & $858^{+30}_{-29}$ 
 & $1260 \pm 100$ & $1130 \pm 30$ \\
Observed 
 & \multicolumn{2}{c}{$583$} 
 & \multicolumn{2}{c}{$864$} 
 & \multicolumn{2}{c}{$1107$} \\
\end{tabular}
\end{ruledtabular}
\end{table*}

Events originating near the top of the detector, dominated by SE emission from the electrodes and radioactivity in the gas region, exhibit narrower S2 widths ($\leq 4~\si{\micro\second}$) and are excluded with an S2 width selection defined by the 99.9\% upper boundary of S1-tagged top events.
We further remove gas-region events with an elevated S2 top-to-total PMT light ratio relative to the $\sim$0.75 expected for bulk S2s \cite{XENONCollaboration:2024bil}.
Finally, events with any S1-like signals within \SI{70}{\micro\second} preceding the S2 are removed \cite{LZ:2025mmr}.
This background is reduced to a negligible level after these selections.

The signal efficiency is obtained from Monte Carlo simulation and is validated using wall, cathode, and $^{37}$Ar calibration L-shell events.
The efficiency as the function of recoil energy is shown in Fig.~\ref{fig:efficiency}.
On average, $\sim$10\% of signal events in the ROI survived data selection.
A systematic relative uncertainty of 11.2\% is assigned, accounting for data selection and fiducial volume uncertainties.

\begin{figure}[t!]
    \centering
    \includegraphics[width=1.0\linewidth]{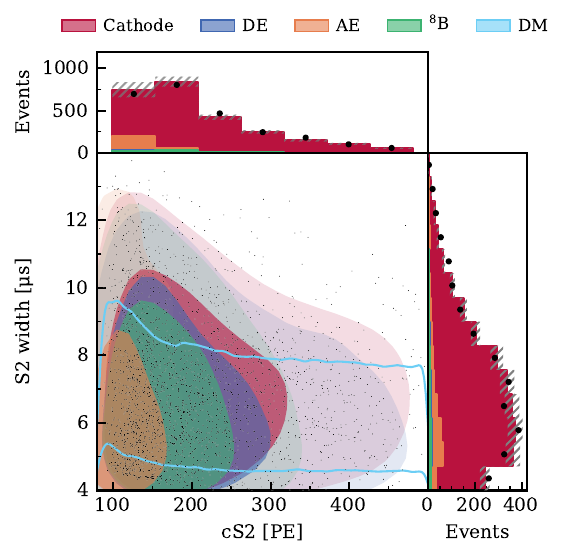}
    \caption{
    \label{fig:science_data}
    Comparison of observed events with expected background components in the science ROI, combining all science runs.
    The description is the same as in Fig.~\ref{fig:background_origin}.
    After applying all data selections, the remaining background has a similar shape to the signal.
    The side projections show the best-fit model and data (black), showing no significant excess.
    The inference is performed in the $\mathrm{cS2}$ dimension.
    }
\end{figure}

The agreement between the background model and science data is shown in two ROIs.
The enlarged ROI shown in Fig. \ref{fig:background_origin} excludes the cathode and DE BDT selections to preserve the features of each background, and is used for background model development.
The science ROI shown in Fig. \ref{fig:science_data} and reported in Table~\ref{tab:table1}, where all selections are applied utilizing the full background rejection power, was blinded and is now used for the DM search.

\itsec{Background modeling}The background model includes cathode, DE, AE, and \beight~\cevns components.
For instrumental backgrounds (cathode, DE, AE), the modeling strategy is developed and validated with $^{220}$Rn \cite{Jorg:2023nvl} and $^{222}$Rn \cite{XENON:2025zff,Jorg:2022tli} calibration data in both the enlarged ROI and the science ROI for each SR.
These calibrations were previously used to model the ER response for the S1–S2 analysis \cite{XENON:2025vwd}.
In the S2-only ROI they contribute negligible ER events, while their increased $\beta$ and $\gamma$ activities at higher energy enhance instrumental backgrounds by increasing large-S2-induced DE rates, raising SE pile-up that generates AE, and accumulation of lead isotopes on the cathode electrode.
Agreement between the background model and the calibration data is observed in reconstructed features \cite{straxen} such as cS2, S2 waveform observables, the CNF score, and event position, validating the model prior to unblinding.
The background validation is further detailed in the End Matter.

For each SR, we build the model in the enlarged ROI in the blinded science data by scaling the normalized shape so that its integral matches the number of events in a background-dominated region (sideband) outside the science ROI.
Residual contamination from other backgrounds in the sideband is subtracted.
The model in the science ROI can be obtained by applying all data selections.
A summary of the background components in the science ROI is provided in Table~\ref{tab:table1}.

For each large S2 $\geq 10^4$~PE in data, we sample the number and features of resulting DE S2s via $P(N,\mathbf{x}\mid\mathbf{y})$ using CNF \cite{Krause:2021ilc}, and salt the simulated DE S2s into the data stream for reprocessing, yielding the normalized DE shape.
A DE-dominated sideband, defined by events rejected by the DE BDT selection and in the region of cS2 [200, 500] PE, and S2 width [23.5, 40] \si{\micro\second} is used for rate normalization.
The resulting DE rates in the science ROI are \SI{1.0(4)}{}, \SI{0.14 \pm 0.027}{}, and \SI{4.4 \pm 0.6} {\eventspertonyear} for SR0, SR1, and SR2, respectively.
The differences in SRs likely arise from SR-dependent DE BDT rejection, while the increase in SR2 is likely due to an event reconstruction algorithm change.

\begin{figure*}[t!]
  \centering
  \begin{minipage}[t]{0.49\textwidth}\vspace{0pt}\centering
    \includegraphics[width=\linewidth]{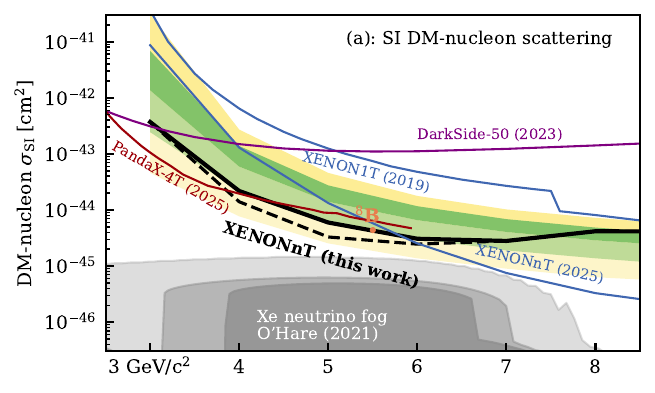}
  \end{minipage}\hfill
  \begin{minipage}[t]{0.49\textwidth}\vspace{0pt}\centering
    \includegraphics[width=\linewidth]{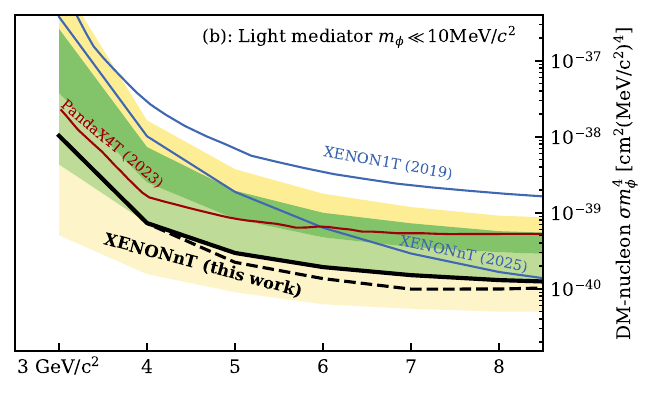}
  \end{minipage}

  \vspace{-0.7em} 

  \begin{minipage}[t]{0.49\textwidth}\vspace{0pt}\centering
    \includegraphics[width=\linewidth]{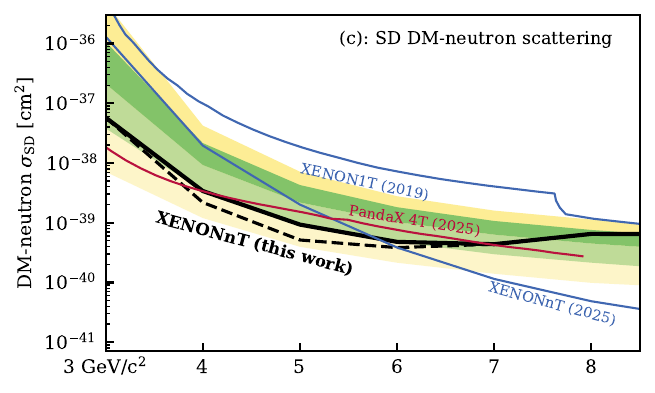}
  \end{minipage}\hfill
  \begin{minipage}[t]{0.49\textwidth}\vspace{0pt}\centering
    \includegraphics[width=\linewidth]{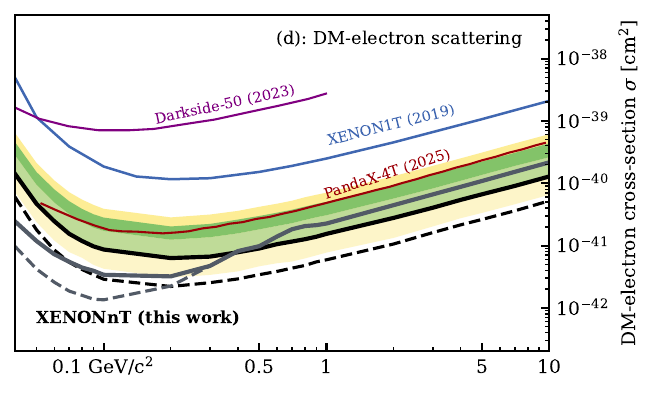}
  \end{minipage}

  \vspace{-0.7em} 

  \begin{minipage}[t]{0.49\textwidth}\vspace{0pt}\centering
    \includegraphics[width=\linewidth]{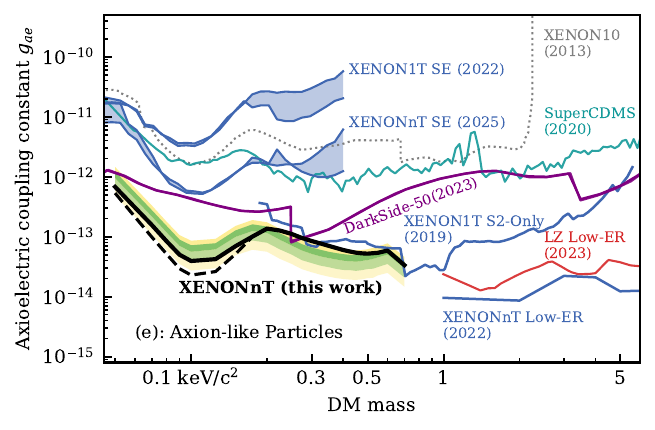}
  \end{minipage}\hfill
  \begin{minipage}[t]{0.49\textwidth}\vspace{0pt}\centering
    \includegraphics[width=\linewidth]{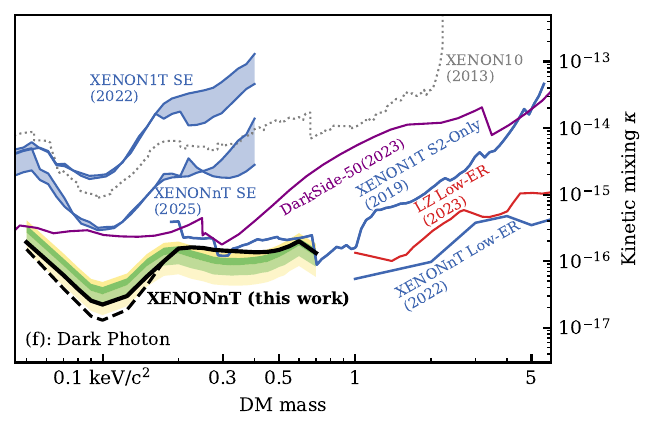}
  \end{minipage}

  \caption{
  The 90\% confidence level upper limits on the DM--particle scattering with 1$\sigma$ (green) and 2$\sigma$ (yellow) sensitivity bands. 
  The black dashed (solid) lines show limits before (after) $-1\sigma$ power-constrained limit (PCL).
  Previous published results from XENON10 \cite{XENON10:2011prx}, XENON1T \cite{XENON:2019gfn}, XENONnT \cite{XENON:2022ltv, XENON:2024hup, XENON:2024znc}, LUX-ZEPLIN \cite{LZ:2023poo}, PandaX \cite{PandaX:2022xqx,Zhang:2025ajc}, DarkSide~\cite{DarkSide-50:2022qzh,DarkSide:2022knj} and SuperCDMS \cite{SuperCDMS:2019jxx} are shown for comparison.
  The neutrino fog region~\cite{OHare:2021utq} in panel (a) is indicated in gray bands for SI DM--nucleon scattering. The \beight~\cevns ~equivalent DM of 5.5 \gevcsq~with cross-section \SI{4.4e-45}{cm^{2}} is shown.
  DM--electron interaction cross section constraints in panel (d) using different atomic ionization factors based on \cite{Essig:2012yx} (\cite{Caddell:2023zsw}) are shown in black (gray) lines, respectively.
  }
  \label{fig:limits}
\end{figure*}

As the reconstruction algorithm merges SEs into S2s regardless of position~\cite{straxen}, AE rejected by the S2 pattern likelihood selection share the same $\mathrm{cS2}$ and S2 width distributions as those leaking into the science ROI, delivering the AE shape.
For rate normalization, we use a sideband defined by S2 area of $\lesssim 4$ extracted electrons ($\sim$120 PE) and S2 widths in [20, 45]~\si{\micro\second}, after subtracting DE leakage.
The resulting AE rates in the science ROI are \SI{65 \pm 11}{} and \SI{45 \pm 3}{\eventspertonyear} for SR0 and SR1, respectively, with no contribution in SR2 due to the higher S2 area threshold.

The cathode background is primarily modeled by waveform simulation, while the $\mathrm{cS2}$ spectrum remains difficult to simulate due to electric field complexity, $^{210}$Pb distribution uncertainty, and wire surface roughness \cite{Linehan:2022hdb}.
Previous studies modeled the S2 area of cathode events using S1–S2 cathode events with progressively smaller S1 signals~\cite{XENON:2019gfn,PandaX:2022xqx,PandaX:2024muv}, but we find that this extrapolation yields an unstable $\mathrm{cS2}$ spectrum that keeps rising toward low energies.
We obtain a sideband $\mathrm{cS2}$ spectrum from a cathode-dominated region defined by large S2 width and low cathode BDT score $\lesssim 0.3$.
After subtracting DE and AE leakage, we rescale the simulated cathode $\mathrm{cS2}$ spectrum to match the sideband $\mathrm{cS2}$ data, using the simulation-driven sideband-to-ROI ratio in each $\mathrm{cS2}$ bins.
Systematic uncertainties from sideband statistics, the rescaling ratio, and leakage of other backgrounds are propagated to the inference.
Cathode radioactivity is the dominant background, with rates in the science ROI of \SI{318 \pm 50}{}, \SI{271 \pm 28}{}, and \SI{310 \pm 23} {\eventspertonyear} for SR0, SR1, and SR2, respectively.

The predicted \beight~\cevns rates in the science ROI are \SI{14 \pm 3}{}, \SI{12 \pm 3}{}, and \SI{8.3 \pm 2.1}{\eventspertonyear} for SR0, SR1, and SR2, respectively, assuming $\sim$3.8\% neutrino flux uncertainty \cite{SNO:2011hxd} and charge yield uncertainty from \cite{XENON:2024kbh}.
The rate is simulated following the same procedure in \cite{XENON:2024hup}.
The rates are higher than those reported in the S1--S2 analysis in \cite{XENON:2024hup} due to the improved signal acceptance.
Radiogenic neutron and electronic recoil backgrounds are expected to contribute fewer than $\mathcal{O}(1)\,\si{\events\per\ton\per\year}$ after data selections and are therefore considered negligible in the analysis.

\itsec{Statistical inference}An extended binned likelihood function is constructed using \textsc{alea} \cite{alea} combining the SR0, SR1, and SR2 data on the cS2 dimension.
Additional dimensions provide little improvement in DM sensitivity in the science ROI.
The likelihood function is expressed as $\mathcal{L}(\sigma,\boldsymbol{\theta}) = \prod_{i=0}^{2} \mathcal{L}_{\mathrm{SR}i}(\sigma,\boldsymbol{\theta}) \prod_m \mathcal{L}^{\mathrm{anc}}_m(\theta_m)$, where $\mathcal{L}_{\mathrm{SR}i}$ is the Poisson likelihood for the cS2 histogram in SR$i$, $\sigma$ is the DM cross section, and $\boldsymbol{\theta}$ collects the nuisance parameters.
The ancillary terms $\mathcal{L}^{\mathrm{anc}}_m$ impose Gaussian constraints on $\theta_m$, including per-SR background rates, signal efficiency nuisances, charge yield shape parameters for $^8$B \cevns and DM signal, and the cathode shape uncertainty from the sideband rescaling.
The nominal background rates and their uncertainties are summarized in Table~\ref{tab:table1}.
The expected sensitivity band is evaluated prior to unblinding by generating toy Monte Carlo datasets \cite{Feldman:1997qc,Baxter:2021pqo}.
We adopt a power-constrained limit (PCL) \cite{Cowan:2011an} with a power threshold of 0.16 to report the final limit.

\itsec{Results}We observe no significant excess in the unblinded science ROI (Table~\ref{tab:table1}, Fig.~\ref{fig:science_data}).
The background-only hypothesis, quantified by a $\chi^2$ goodness-of-fit test on the cS2 dimension, is compatible with the SR0 and SR1 data, with p-values of 0.997 and 0.552, while SR2 shows a mild tension, with p-value of 0.013 due to a downward fluctuation. 
A discrepancy between data-driven and simulation-derived efficiency was observed in SR2 for $\mathrm{cS2}<200$~PE after unblinding, caused by the temporal isolation selection designed to suppress AE by requiring no nearby S2s.
Since the salting method cannot simulate ambient S2s generated by DM events themselves, this selection was less effective on salted waveforms than on real data, leading to an overestimation of signal efficiency. Therefore, the selection was relaxed to consider only ambient S2s preceding DM candidates.
This conservative adjustment has negligible impact on the final limits.

We set 90\% confidence level upper limits with expected sensitivity bands shown in Fig.~\ref{fig:limits}.
For SI DM–nucleon scattering we exclude cross sections above $6.0\times10^{-45}\,\text{cm}^2$ at $m_\chi=5~\gevcsq$, probing new parameter space in DM masses $[4,6]~\gevcsq$.
We also exclude new parameter spaces for spin-dependent DM--neutron scattering and light-mediator scenarios.
Low-energy ER models benefit most in the S2-only analysis, as their S1 falls below the S1–S2 threshold while the S2 remains detectable, excluding DM–electron scattering $\sigma$ above $6.3\times10^{-42}\,\text{cm}^2$ at $m_\chi=0.2~\gevcsq$.
We exclude more parameter spaces for bosonic DM candidates such as axion-like particles and dark photons, at $g_{ae}=3.9\times10^{-14}$ and $\kappa=2.2\times10^{-17}$ at $m_\chi=0.1~\mathrm{keV/c^2}$, respectively.

\itsec{Discussion}Accurate modeling of instrumental backgrounds is crucial for DM searches near the detector threshold \cite{Baxter:2025odk}.
This work establishes a systematic data selection and background modeling methodology for ionization-only searches, using novel machine learning methods benefiting from accurate detector modeling.
We find that radioactivity from the cathode electrode dominates the residual background, underscoring the importance of electrode design and material purity to enhance the sensitivity to light DM and solar neutrinos for next-generation DM detectors \cite{XLZD:2024nsu,PANDA-X:2024dlo}.
The study also indicates that isolated S2s that contribute to accidental coincidences in S1–S2 analyses \cite{XENON:2024hup,XENON:2024ijk,LZ:2025igz} may originate primarily from the cathode electrode.
Future work will focus on developing a simulation-driven cathode background model and a complete background model down to the single electron level.


\itsec{Acknowledgements}We gratefully acknowledge support from the National Science Foundation, Swiss National Science Foundation, German Ministry for Education and Research, Max Planck Gesellschaft, Deutsche Forschungsgemeinschaft, Helmholtz Association, Dutch Research Council (NWO), Fundacao para a Ciencia e Tecnologia, Weizmann Institute of Science, Binational Science Foundation, Région des Pays de la Loire, Knut and Alice Wallenberg Foundation, Kavli Foundation, JSPS Kakenhi, JST FOREST Program, and ERAN in Japan, Tsinghua University Initiative Scientific Research Program, National Natural Science Foundation of China, Ministry of Education of China, DIM-ACAV+ Région Ile-de-France, and Istituto Nazionale di Fisica Nucleare. This project has received funding/support from the European Union’s Horizon 2020 and Horizon Europe research and innovation programs under the Marie Skłodowska-Curie grant agreements No 860881-HIDDeN and No 101081465-AUFRANDE.

We gratefully acknowledge support for providing computing and data-processing resources of the Open Science Pool and the European Grid Initiative, at the following computing centers: the CNRS/IN2P3 (Lyon - France), the Dutch national e-infrastructure with the support of SURF Cooperative, the Nikhef Data-Processing Facility (Amsterdam - Netherlands), the INFN-CNAF (Bologna - Italy), the San Diego Supercomputer Center (San Diego - USA) and the Enrico Fermi Institute (Chicago - USA). We acknowledge the support of the Research Computing Center (RCC) at The University of Chicago for providing computing resources for data analysis.

We thank the INFN Laboratori Nazionali del Gran Sasso for hosting and supporting the XENON project.\\

\begin{figure}[t!]
    \centering
    \includegraphics[width=1.0\linewidth]{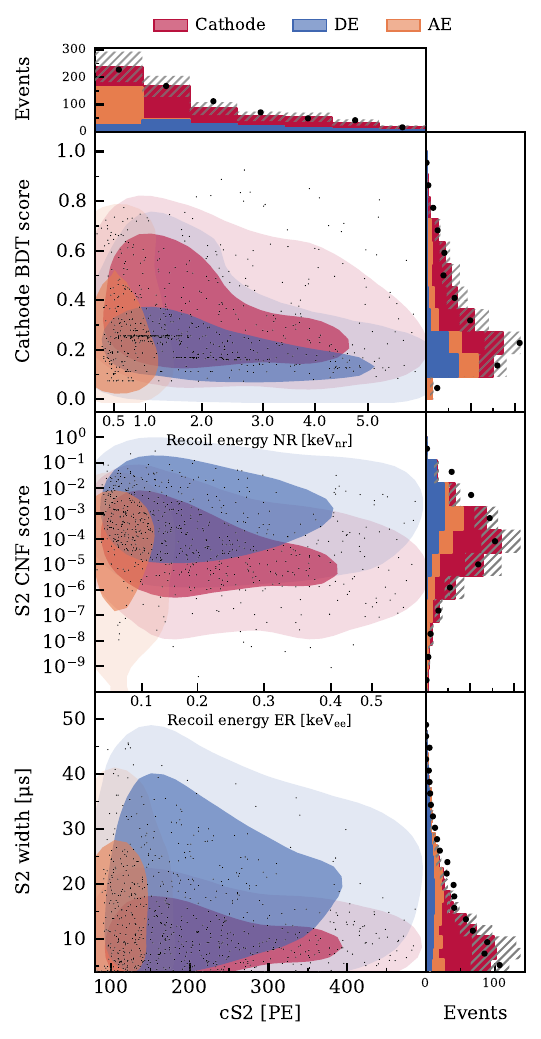}
    \caption{Comparison between observed events and the background model in the enlarged ROI for the $^{220}$Rn calibration, combining SR0, SR1, and SR2.
    Matching between data and model is observed in dimensions except the S2 CNF score, potentially due to the substantially enhanced $\gamma$-ray activity in the $^{220}$Rn calibration, as the CNF model is trained using the science data.
    The S2 CNF score is further validated using the $^{222}$Rn calibration, where the high-energy event rate is lower.
    The \beight~\cevns contribution is negligible because of the limited live time.
    \label{fig:calibration_data_rn220}
    }
\end{figure}

\begin{figure}[t!]
    \centering
    \includegraphics[width=1.0\linewidth]{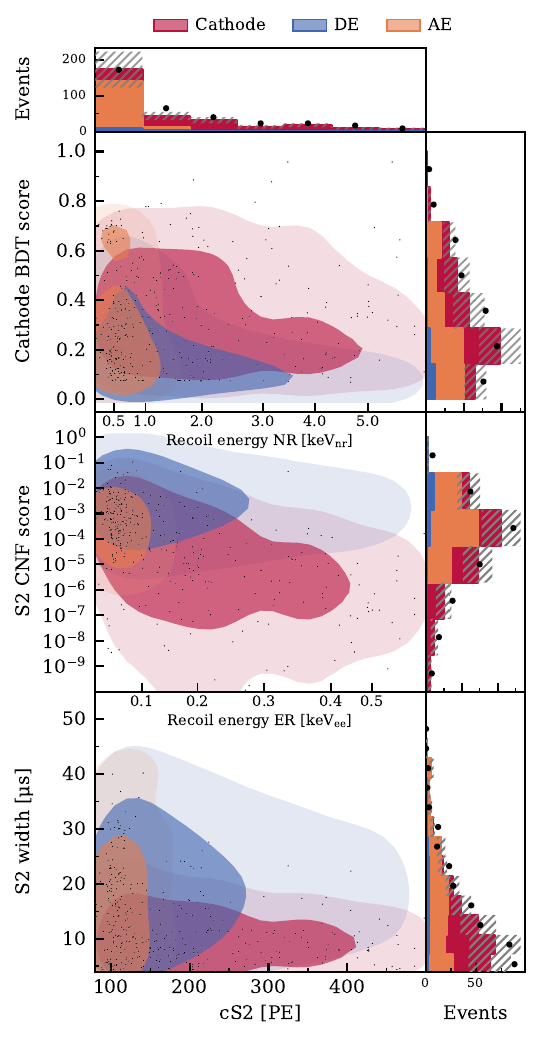}
    \caption{Comparison between observed events and the background model in the enlarged ROI for the $^{222}$Rn calibration in SR1.
    The $^{222}$Rn injection activity is lower than that of the $^{220}$Rn calibration, yielding a background rate and cathode background isotope origin more similar to those of the science data.
    Good agreement between data and model is observed in all four dimensions, including the S2 CNF score.
    \label{fig:calibration_data_rn222}
    }
\end{figure}

\itsec{Appendix A: Background validation}Validating the completeness of the background model is essential for a DM search.
The calibration datasets using radioactive sources provide background-dominated samples; the short livetimes make contributions from rare physics processes, such as DM and solar neutrino \cevns interactions negligible, while their high event rates enhance the instrumental backgrounds relevant to this analysis.
The instrumental background model is validated using the non-blinded $^{220}$Rn (Fig.~\ref{fig:calibration_data_rn220}) and $^{222}$Rn (Fig.~\ref{fig:calibration_data_rn222}) calibration datasets prior to the unblinding of the science data. 
A temporary enhancement of instrumental backgrounds is observed in both calibrations, due to the increased overall event activity in the TPC associated with the intense radon daughter decays and $\gamma$-ray activity.

\begin{figure}[!t]
    \centering
    \includegraphics[width=1\linewidth]{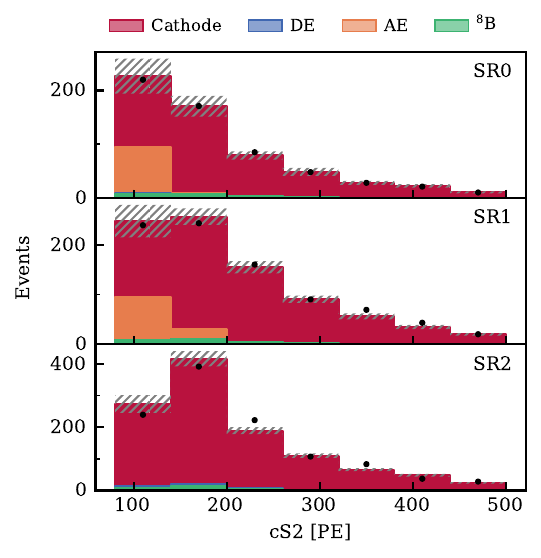}
    \caption{Observed events and the best-fit background model comparison in the cS2 dimension for SR0, SR1, and SR2.
    The spectrum in SR2 differs from previous SRs due to the increased S2 reconstruction threshold of 120 PE.}
    \label{fig:data_per_sr}
\end{figure}

\itsec{Appendix B: Observed events per SR}The three SRs are characterized by varying impurity levels, notably the different rates of DE and AE backgrounds, as well as the increased S2 threshold in SR2. 
To demonstrate the validity of the background model across the changing operational phases, Fig.~\ref{fig:data_per_sr} presents a comparison between the unblinded science data and the best-fit background model in the $\mathrm{cS2}$ inference dimension for each SR.
The observed events consistently match the best-fit background distributions, confirming the robustness of the background modeling.

\bibliography{main}
\nocite{data_release}

\end{document}